# The Breakdown of Dynamic Scaling and Intermittency in a Cascade Model of Turbulence

Omri Gat, Itamar Procaccia and Reuven Zeitak

*Department of Chemical Physics*
*The Weizmann Institute of Science*
*76100 Rehovot, Israel*

**Abstract**

We present an analytic and numerical analysis of the Gledzer-Ohkitani-Yamada (GOY) cascade model for turbulence. We concentrate on the dynamic correlations, and demonstrate both numerically and analytically, using resummed perturbation theory, that the correlations do not follow a dynamic scaling ansatz. The basic reason for this is the existence of a second quadratic invariant, in addition to energy. This implies the breakdown of the Kolmogorov type scaling law, in a manner different from the conventional mechanisms proposed for Navier-Stokes intermittency. By modifying the model equation so as to eliminate the spurious invariant, we recover to good accuracy, both dynamic scaling and the Kolmogorov exponents. We conclude that intermittency in the GOY model may be attributed to the effects of the spurious invariant which does not exist in the 3-dimensional Navier-Stokes flow.

# 1 Introduction

A class of models of hydrodynamic turbulence that has recently attracted some interest are the so called 'shell' or Gledzer-Ohkitani-Yamada (GOY) [1][2] models. These models exhibit numerically a clear deviation from the expected scaling predicted by an application of Kolmogorov like dimensional analysis. In addition, high moments of the variables exhibit *multiscaling*. It has been speculated that the mechanism for generation of multiscaling in these models may be related to the possibility of multiscaling in Navier-Stokes turbulence. Intermittency in hydrodynamic turbulence is usually attributed to large spatial fluctuations in the local energy dissipation $\epsilon$. In the shell models there is no space coordinate, and therefore it is unclear how this mechanism could be applied to shell models. However, reference [2] considers the connection between dissipation fluctuations in time and intermittency in the structure function via a 'Taylor hypothesis' type argument.

We demonstrate in this paper that the "intermittency" observed in the GOY models is largely and perhaps completely due the existence of a second quadratic invariant in addition to energy [3] [4]. This quadratic invariant does not have an analog in Navier-Stokes equations, and is not even positive definite, but still its existence will be shown to have important effects on the dynamics. This claim will be supported by comparison with a modified version of the GOY model that lacks the second invariant. In particular the original GOY does not possess dynamic scaling, while the modified version obeys dynamic scaling.

The paper is organized as follows; Section 2 defines the model and details some of its important symmetries and conserved quantities. In section 3 we present our numerical results and demonstrate the absence of dynamic scaling in the GOY model. In section 4 we examine the direct interaction approximation (DIA) equations for the GOY model and conclude that dynamic scaling is indeed inconsistent with the DIA due to the existence of the second invariant. In section 5 we consider the modified GOY and show that dynamic scaling and the Kolmogorov exponents are restored. Section 6 offers some conclusions regarding the relations between intermittency in shell models and Navier-Stokes turbulence.

# 2 General properties of the model

## 2.1 Model definition

The GOY model reads (in its general form)[1],

$$(\partial_t + \nu\, k_n^2)u_n = i[a\, k_n\, u_{n+1}u_{n+2} + b\, k_{n-1}\, u_{n-1}u_{n+1} + c\, k_{n-2}\, u_{n-1}u_{n-2}]^* + f_n, \quad (1)$$

where $u_n$, $n \geq 0$ is the complex dynamical variable of shell $n$, with wave number $k_n = k_0 q^n$. $q > 1$ is the shell spacing, $\nu$ is the 'viscosity', $f_n$ is the external forcing of shell $n$, and $a$, $b$ and $c$ are real coupling constants whose sum is zero. The number of shells may in principle extend to infinity but in practice the presence of the viscosity term acts as a cut-off. This allows us to truncate the system of equations after a finite number of levels without changing the dynamics.

## 2.2 A phase symmetry

The unforced Navier Stokes equation possesses translational invariance. In $k$ space, this implies an invariance of the equation with respect to the transformation $\mathbf{v_k} \to e^{i\mathbf{k}\cdot\mathbf{r}_0}\mathbf{v_k}$ of a solution $\mathbf{v_k}$. This symmetry implies that the Fourier components of the velocity field are delta correlated in $k$, e.g.

$$\langle \mathbf{v_k}\mathbf{v_q} \rangle = F(\mathbf{k})\delta(\mathbf{k}+\mathbf{q}). \tag{2}$$

This follows from the assumption that the symmetry of the equation is not broken by the forcing. The implication of this relationship is that the pair correlators and the Green's functions are all diagonal in $k$, in other words, the vertex conserves momentum.

The translational invariance of the Navier-Stokes is not retained in the GOY model. However, there exists an analog of this symmetry which leaves the GOY model equations invariant. This is the transformation

$$u_n \to e^{i\theta}u_n \tag{3}$$
$$u_{n+1} \to e^{i(\phi-\theta)}u_{n+1} \tag{4}$$
$$u_{n+2} \to e^{-i\phi}u_{n+2} \tag{5}$$

Note that $\theta$ and $\phi$ are arbitrary global phases. This transformation adds the same phase to shell variables that are separated by a distance of three shells. A reasoning similar to that of the Navier-Stokes case leads to the conclusion that

$$\langle u_n u^*_{n+m} \rangle = 0 \quad \text{if } m \text{ is not divisible by 3.} \tag{6}$$

This symmetry also implies that the shell variables have a zero mean. Similar relations may be obtained for higher order correlations.

## 2.3 Quadratic conserved quantities

A crucial ingredient of the Kolmogorov argument is the existence of one (quadratic) invariant for the Euler equations, energy. For the Navier-Stokes equations The energy is dissipated at the smallest scales and is injected at the largest scales. Kolmogorov assumed the existence of an energy flux that is scale independent in the inertial range. Motivated by this, it is useful to analyze the GOY model in terms of quadratic invariants.

We are interested in conserved quantities of the form

$$C = \sum A_n |u_n|^2 \tag{7}$$

and demand that in the limit of no forcing or viscosity $\partial_t C \equiv 0$, independent of $u_n$. Hence, inserting in equation (1) we obtain a system of equations

$$A_n a + A_{n+1} b + A_{n+2} c = 0. \tag{8}$$

This system may be viewed as linear recursion relation for $A_{n+2}$ in terms of $A_{n+1}, A_n$. This means that there are only two linearly independent solutions for the $A$'s. As we demand energy conservation, one of these solutions is $A_n \equiv 1$ and consequently $a + b + c = 0$. We look for the second independent solution in the form $A_n = A^n$. We may solve for $A$ yielding $A = a/c$. With the standard choice of parameters $a = 1, b = c = -1/2$ we obtain $A_n = (-2)^n = (-)^n k_n$. This new conserved quantity $L = \sum (a/c)^n |u_n|^2$ is spurious in the sense that it has a dependence on the parameters of the model. In the usual choice of parameters (see eq (12) below) this invariant is not positive definite, however this does not prevent it from having an important dynamical effect [3][4] [5] (see below). For a particular set of parameters, it is possible to interpret $L$ as the analog of helicity which is conserved in the Euler equations [3]. However the existence of this invariant was not considered as a requirement in the original definition of the model [1].

## 3 Numerical examination of scaling properties

In previous works [2][6] it has been found that the simultaneous correlators of the shell variables obey scaling laws, such as

$$\langle |u_n|^q \rangle \sim k_n^{-\zeta_q}, \tag{9}$$

for shell numbers inside the inertial range. $\zeta_q$ was found to have a non-linear dependence on $q$. Such a behaviour is termed 'multiscaling'. The inertial range is defined as the shells in which both the dissipative and the forcing terms are small with respect to the non-linear term.

The usual Kolmogorov picture implies also that the different-time correlators should obey a dynamic scaling relation of the type

$$\langle u_n^*(t) u_n(t+\tau) \rangle \sim k_n^{\zeta_2} \hat{f}(\tau k_n^z). \tag{10}$$

$\hat{f}$ is a universal scaling function independent of the specific shell. This scaling relation becomes in the frequency domain

$$\langle |u_n(\omega)|^2 \rangle \sim k_n^y f(\omega/k_n^z), \tag{11}$$

where $y + z = \zeta_2$. Kolmogorov scaling predicts that $y = -4/3$ and $z = 2/3$. Although intermittency may change the values of these exponents, one would still expect naively the dynamic scaling hypothesis to hold.

In order to test this hypothesis we integrated numerically the GOY equations and considered the frequency power spectra of the shell variables. Following the usual procedure we chose $f_3 = (1+i)5 \times 10^{-3}$ and all other $f_n$'s to be zero. In the numerical integration of the equations we used the following parameters:

$$a = 1, \qquad b = -0.5 \qquad c = -0.5$$
$$k_0 = 1/(32\sqrt{2}) \quad \nu = 6 \times 10^{-7} \quad q = \sqrt{2}. \qquad (12)$$

There were 43 shells (running from 0 to 42) in the model which corresponds to roughly 21 level in the $q = 2$ models. The equation were integrated using a 4th-5th order Runge-Kutta adaptive time step method. The integration time was $6.6 \times 10^4$ which corresponds to approximately 1500 forcing scale turnover times.

The power spectra were calculated as follows: A data string of 65536 equally time-spaced samples, calculated by linear interpolation of the values obtained from the integrator, were fast Fourier transformed with a Hanning window, and $|u_n(\omega)|^2$ was calculated. This procedure was repeated many times and the values of $|u_n(\omega)|^2$ were averaged.

The spacing between the samples was chosen to be somewhat smaller than the characteristic time scale for the highest shell (the 37th) which was analyzed. The size of the sample was sufficient to capture the whole dynamic range of the power spectrum except for the lowest shells.

An attempt to collapse the data of several power spectra is shown in figure 1. The power spectra are rescaled according to the typical height and width of each power spectrum as follows:

$$f_n^{(r)}(w) = \frac{1}{h_n} f_n(w \omega_n), \qquad (13)$$

where

$$\omega_n = \left( \frac{\int d\omega \omega^2 f_n(\omega)}{\int d\omega f_n(\omega)} \right)^{1/2}, \qquad h_n = \frac{\int d\omega f_n(\omega)}{\omega_n}; \qquad (14)$$

$w$ is the dimensionless frequency $\omega/\omega_n$.

The three power spectra shown in figure 1 demonstrate that the rescaling performed in (14) yields a poor data collapse. There is a general trend for the spectra of higher shells to be flatter than lower shells at the same rescaled frequencies. We thus conclude that dynamic scaling is incompatible with our numerical results. In the next section we will analyze the GOY model analytically and show explicitly where dynamic scaling fails.

We also measured the cross correlations between shells. Recall that the symmetry (3) implies a vanishing of cross correlations that have a non zero modulo three difference between them. This effect was clearly observed in our simulation. Cross correlations of shells differing by a multiple of three were larger by two orders of magnitude compared to cross correlations of nearby shells without this property. The normalized cross correlations ($\langle u_n u_m^* \rangle / \sqrt{\langle |u_n|^2 \rangle \langle |u_m|^2 \rangle}$) decay as $|n - m|$ increases (by multiples of three) but seem to converge to a non zero value when one of the shells is in the dissipation range (larger than 33).

Note that the period three oscillations observed by many authors [3][6] are a direct result of aliasing of the effect of the forcing into the inertial range by the period three level to level correlations. They should be interpreted as a peculiar artifact of the GOY model and have no relation to 'lacunarity' in real turbulence. This is clearly shown by the fact that the period three behavior is independent of the value of the shell separation and will persist even in the limit of shell separation going to zero [7] [8]. On the other hand, the period 2 oscillations in the structure functions near the top of the inertial, are genuine dynamical effects, and are due to the fact that $L$ is not positive definite [3].

In addition we measured the scaling of the structure functions $\langle |u_n|^q \rangle$. These quantities are known to exhibit *multiscaling* [2] and we have recovered this behavior. In fact, as we are using shells separated by $\sqrt{2}$, we found stronger effects of intermittency (similar to what was observed in [6] ) .

## 4 Direct interaction approximation for the GOY model

As we have seen in the previous section, dynamic scaling fails to explain the shape of the frequency power spectrum. In order to examine this effect analytically we performed a resummed perturbation expansion for the GOY model. Basically we have derived the analog of the DIA equations [9] for the GOY model. The DIA approximation is known to fail for the Navier-Stokes equations due to the failure to respect Galilean invariance, or not taking into account the effect of large scale advection in a proper way [10]. The GOY model forms a truncated set of equations which are only coupled locally, hence conventional wisdom (which we will have no reason to doubt in what follows) indicates that the DIA should be a good approximation for the GOY model.

As we have noted in section 2.2 the phase symmetries (3) induce the vanishing of some cross terms in the correlation functions and in the Green functions, however, not all of the cross terms vanish. This implies that, as opposed to the Navier-Stokes case, the GOY model DIA equations are formulated in terms of functions that are not diagonal in the shell $n$ index, namely

$$\begin{aligned} G(n, m, \omega) \delta(\omega + \omega') &\equiv \langle \delta u_n(\omega) / \delta f_m(\omega') \rangle, \\ U(n, m, \omega) \delta(\omega + \omega') &\equiv \langle u_n(\omega)^* u_m(\omega') \rangle. \end{aligned} \quad (15)$$

The non-diagonality makes for a slightly more complicated form for the DIA equations. They now contain expressions for off diagonal quantities as well as diagonal ones. It also complicates the discussion of the convergence properties of the integrals and the sums. Nevertheless, one obtains the self consistent Dyson-Wyld integral equations,

$$G(n,m,\omega) = G^0(n,\omega)(\delta_{n,m} + \sum_l \Sigma(n,l;\omega)G(l,m;\omega)) \quad (16)$$

$$U(n,m;\omega) = \sum_{k,l} G^*(n,k;\omega)(\langle f_k^* f_l \rangle + \Phi(k,l;\omega))G(l,m;\omega), \quad (17)$$

and the DIA for the mass operators

$$\Sigma^{(2)}(n,m;\omega) = \int d\omega' k_n k_m \sum_{ii'} \sum_{jj'} c_{ii'} c_{jj'} q^{-j'} U(n+i, m-j, -\omega-\omega')$$
$$\times G(n+i', m-j', \omega') \quad (18)$$

$$\Phi^{(2)}(n,m;\omega) = \frac{1}{2} \int d\omega' k_n k_m \sum_{ii'} \sum_{jj'} c_{ii'} c_{jj'} U(n+i, m+j, -\omega-\omega')$$
$$\times U(n+i', m+j', \omega'), \quad (19)$$

where the coupling coefficients are

$$c_{1,2} = c_{2,1} = a, \qquad c_{1,-1} = c_{-1,1} = b, \qquad c_{-2,-1} = c_{-1,-2} = c.$$

Although in principle one should solve these equations to find the behavior of the correlation function, one usually tries a scaling ansatz in order to extract information about the scaling behavior of the correlation function $U$. Not withstanding our knowledge based on numerics that dynamic scaling fails, it is instructive to try a dynamic scaling form for $G$ and $U$ in order to see why it fails. We assume

$$G(n,m,\omega) = (k_n k_m)^{-z/2} g(\tfrac{\omega}{k_n^z}, k_m/k_n).$$
$$U(n,m,\omega) = (k_n k_m)^{y/2} f(\tfrac{\omega}{k_n^z}, k_m/k_n). \quad (20)$$

where $y$ and $z$ are scaling exponents to be self consistently determined. If we assume naively that all sums and integrals converge in the limit of infinitesimal dissipation[11] ($\nu \to 0$) we find that the DIA equations for $F$ and $G$ both yield the same exponent relation

$$z - y = 2 \quad (21)$$

Thus up to this point dynamic scaling gives us a one parameter family of solutions that is self consistent. In order to obtain a second exponent relation it is usual to use the conservation of energy in the inviscid limit (Euler equation) to deduce a conservation of flux in the inertial range (this conservation may be considered as the definition of the inertial range). The energy flux through shell $n$, is defined in the GOY model as [6] [3]

$$\epsilon_n = -\Im m \langle -c k_{n-1} u_{n-1} u_n u_{n+1} + a k_n u_n u_{n+1} u_{n+2} \rangle. \quad (22)$$

Assuming dynamic scaling, the lowest order terms in the expansion for the 3-point correlators in (22) scale as $k_n^{1+z+2y}$, thus constancy of the the flux (tested numerically by [6]) implies that

$$z + 2y = -2, \qquad (23)$$

which with (21) gives $z = 2/3$, $y = -4/3$, which is what a straightforward application of Kolmogorov type arguments gives [1].

Renormalized perturbation expansions in which Galilean invariance is preserved [12] [13] predict Kolmogorov scaling for Navier-Stokes turbulence, including dynamic scaling for the correlation functions and Green's functions. However, experiments indicate that the scaling of higher order structure functions differs from the K41 predictions. How to understand these deviations and how to reconcile these with the results of renormalized perturbation theories are a subject of active study [14][15]. At any rate, none of the present explanations resorts to assuming the failure of dynamic scaling as a cause. Indeed, in Navier-Stokes turbulence there seems to be no reason to doubt the existence of dynamic scaling.

The GOY model is different in this respect. In addition to relationship (22) we have an additional conserved quadratic quantity $L$ defined in 2.2 . This induces an additional conserved flux

$$\ell_n = -\Im m \langle (a/c)(-k_{n-1} u_{n-1} u_n u_{n+1} + k_n u_n u_{n+1} u_{n+2}) \rangle. \qquad (24)$$

which yields the exponent relation

$$z + 2y = -2 - \frac{\log(a/c)}{\log q} \qquad (25)$$

in the same manner as (23) was obtained. The RHS of this scaling relation is not real in general. Such a solution is unacceptable, since the simultaneous pair correlation function must be positive. A more general scaling ansatz that includes a periodic function of $n + m$ can solve this problem, and then only the real part of eq. (25) should be taken. However this scaling

$$z + 2y = 2 + \frac{\log |a/c|}{\log q} \qquad (26)$$

contradicts the scaling relation (23) derived from energy flux conservation. Indeed, the energy flux conservation is violated when the original scaling ansatz is multiplied by a non-constant periodic function of $n + m$.

There are two ways to solve the contradiction. Either one of the fluxes vanishes in the inertial range (similar to the case of 2 dimensional turbulence) or we must reject the assumption of dynamic scaling. If one of the fluxes is small in some sense compared to the other one may treat it as yielding a correction to dynamic scaling superimposed on top of the larger flux's contribution. Our numerics do not support this case but we cannot rule out the possibility that intermittency disappears asymptotically in the number of shells. The existence of two non vanishing fluxes in the inertial range is incompatible with dynamic scaling. A numerical check of the flux $\ell$ indeed demonstrates that it doesn't vanish. When only one level is forced, the two fluxes are linearly dependent [3], and therefore must be non-zero simultaneously. We identify this fact as the reason for the failure of dynamic scaling observed numerically in section 3.

## 5  A shell model with a single invariant

Following the reasoning of the previous section, modifying the GOY model so as to remove the spurious $L$ invariant should restore the possibility of a dynamic scaling solution for the DIA equations. A dynamic scaling solution of the DIA entails the Kolmogorov scaling for simultaneous two point correlations. If the DIA is a good approximation then we cannot find a dynamic scaling solution of the GOY with intermittency. This prediction can be tested in a model that obeys dynamic scaling, thus we must construct a model with only one conserved flux.

A simple way of eliminating the $L$ invariant is by coupling the shell variables to higher and lower shells. A generalization of the GOY interaction term that couples to more modes reads

$$i(\sum_{i<j} a_{ij}k_n u_{n+i}u_{n+j} + b_{ij}k_{n-i}u_{n-i}u_{n+j-i} + c_{ij}k_{n-j}u_{n-j}u_{n+i-j})^*. \qquad (27)$$

The sum runs on a finite set of pairs (the original GOY is the special case where there is only one pair, namely $\{1,2\}$). Energy conservation in the inviscid unforced case demands that

$$a_{ij} + b_{ij} + c_{ij} = 0. \qquad (28)$$

The condition (8) whose solutions are the quadratic invariants of the system is generalized in this case to

$$A_n a_{ij} + A_{n+i} b_{ij} + A_{n+j} c_{ij} = 0, \qquad \text{for every pair } \{ij\}. \qquad (29)$$

The condition (28) ensures that (29) is satisfied by $A_n = const$. The condition (29) is clearly more restrictive than the original condition (8), and we will show in the following an explicit example in which energy is the only conserved quadratic invariant.

In principle the pairs $\{ij\}$ may chosen arbitrarily. However, if we want to ensure the vanishing of $\langle u_n \rangle$ directly from the symmetry of the equations, we need the phase symmetry (3). Thus we are interested in including only pairs which preserve (3). This set includes all the pairs in which $i+j$ is divisible by 3, and neither $i$ nor $j$ is divisible by 3.

We now have quite a large parameter space at our disposal, of which we chose the next most local allowable pair which is $\{2,4\}$ in addition to $\{1,2\}$.

Returning to equation (29) with the specific choice of pairs $\{2,4\},\{1,2\}$ we see that there are two independent ways to obtain $A_{n+4}$ from $A_n, A_{n+1}$. One way, by iterating the equation for the $\{1,2\}$ pair twice, and another way is by iterating it once to get an expression for $A_{n+2}$ and inserting this into the equation for the $\{2,4\}$ pair. As both expressions must be equal, we find a mapping that gives $A_{n+1}$ as a function of $A_n$. As we know that $A_n = A_{n+1} = A_{n+2} = \cdots$ is a solution due to (28), we conclude that there are no more independent quadratic invariants for this coupling choice.

Our choice of couplings means that there is a 3 dimensional parameter space, since each pair involves 2 independent coupling constants and in addition a global scale factor can be absorbed in a redefinition of the $u_n$ variables. As observed by [16], the original GOY doesn't exhibit chaotic behaviour in parts of the parameter space; instead there can be a stable fixed point, limit cycle, or quasi-periodic motion. This phenomenon persists also in the enlarged parameter space [8]. We expect dynamic scaling to appear for parameter values for which there are chaotic solutions. After a qualitative mapping of the parameter space we found a region which has the desired property. Typical parameter values in this region are

$$a_{12} = 1 \quad b_{12} = -0.5 \quad c_{12} = -0.5$$
$$a_{24} = 5 \quad b_{24} = -2.5 \quad c_{24} = -2.5 \tag{30}$$

This choice of parameters is essentially two original GOY-like couplings, the second of which is 5 times stronger than the first. The fact that the $\{2,4\}$ coupling is much stronger than the $\{1,2\}$ coupling raises a problem that if only one level is forced, the shells with the same parity as the forced shell will have larger amplitudes than neighboring shells with opposite parity. This problem is dealt with by forcing shell 2 by $f_2 = (1+i)10^{-2}$ in addition to the usual forcing of shell 3, but even then there remain some period 2 oscillations in the structure functions. Other numerical parameters remain unchanged, including the simulation duration.

We have measured the same quantities as for the original GOY model (3). The data collapse for the frequency domain power spectra is shown in figure 2. The fit is clearly superior to the fit accomplished in figure 1. Also, the empirical rescaling parameters $h_n$ and $\omega_n$ defined in 14 can be fitted well to power laws

$$\omega_n \sim k_n^{.67 \pm .01} \tag{31}$$
$$h_n \sim k_n^{-1.35 \pm .01} \tag{32}$$

Which are compatible with the dynamic scaling prediction and are close to the Kolmogorov exponents $2/3, -4/3$. As another check we have measured higher order structure functions $\langle |u_n|^q \rangle$ and compared their scaling exponents to those of the original GOY and to Kolmogorov scaling. Table 1 contains this comparison. Note that higher moments still do not conform to the Kolmogorov exponents but the deviation is much smaller.

The deviation may be attributed to many sources, such as finite system effects (corrections to scaling), a persistence of some memory of the extra invariant and, of course, real intermittency, indicating a failure of the perturbation scheme on the level of the DIA. It is beyond the scope of this paper to decide which of the above effects is responsible for the deviations from Kolmogorov scaling. Our purpose is to stress that these effects are much smaller than the ones observed in the original GOY model. We attribute the large deviations in the original GOY to the lack of dynamic scaling that is induced by the existence of a second flux in the inertial range.

# 6 Conclusions

Numerical evidence shows that the solutions of the GOY model do not have dynamic scaling. In addition, the DIA equations do not have solutions which exhibit dynamic scaling, due to the existence of two non vanishing constant fluxes in the inertial range. The derivation of Kolmogorov scaling for the structure functions via perturbation methods relies on the existence of dynamic scaling. Therefore, there is no evidence for a contradiction between perturbation theory and multiscaling in the GOY model.

This situation is unlike the case of Navier-Stokes turbulence. Experimentally, high order structure functions deviate from the predicted Kolmogorov exponents [17]. However, we have no good reason to doubt the scale invariance in time and space of turbulent motion. Helicity, or other conserved quantities which in principle give rise to conserved fluxes, are zero in isotropic homogeneous turbulence, and are considered small in theories of turbulence. A dynamic scaling ansatz in the perturbation theory is then consistent and yields the Kolmogorov scaling uniquely [12][18]. It should be stressed that dynamic scaling has never been proved from first principles, and its failure may be an interesting alternative mechanism for explaining some aspects of turbulent phenomenology. The results of this paper motivate us to propose that dynamic scaling should be tested directly in experiments in turbulence. A major deviation from dynamic scaling would imply that a reconsideration of the current working assumptions in theories of turbulence is needed. Whether or not the final resolution of intermittency in turbulence will rely on presently applied techniques or whether non-perturbative effects should be considered is still not clear.

The reason for the difference between the Navier-Stokes intermittency and the intermittency in the GOY model is the existence of a second quadratic invariant. Intermittency in a model in which the second invariant is absent may have an origin closer to the Navier-Stokes type intermittency. However, our results seem to indicate that a modified GOY model without the second invariant has only small remnants of intermittency. Therefore, in our opinion, the relevance of multiscaling in the GOY to the dynamics of Navier-Stokes is questionable.

## Acknowledgments

We would like to acknowledge discussions with Gregory Falkovich, Gilad Goren, Detlef Lohse, Victor Lvov and Zeev Olami. This work was supported in part by the German-Israeli Foundation. We also thank Detlef Lohse for communicating reference [3] to us prior to publication.

| $n$ | 1 | 2 | 3 | 4 | 5 | 6 |
|---|---|---|---|---|---|---|
| Original GOY ($q=\sqrt{2}$) | .42 | .78 | 1.05 | 1.20 | 1.29 | 1.35 |
| Modified GOY ($q=\sqrt{2}$) | .36 | .69 | 1.00 | 1.28 | 1.53 | 1.81 |
| Original GOY ($q=2$) | .39 | .75 | 1.07 | 1.36 | 1.64 | 1.87 |
| Modified GOY ($q=2$) | .35 | .69 | 1.03 | 1.35 | 1.65 | 1.92 |
| K41 | .33 | .67 | 1.00 | 1.33 | 1.67 | 2.00 |

Table 1: A comparison between the scaling exponents $\zeta_n$ of the structure functions in a few models and Kolmogorv scaling. The errors in the scaling exponents are approximately $\pm.01$, except for the modified model with $q=2$ where they are $\pm.03$, due to large fluctuations in the structure functions.

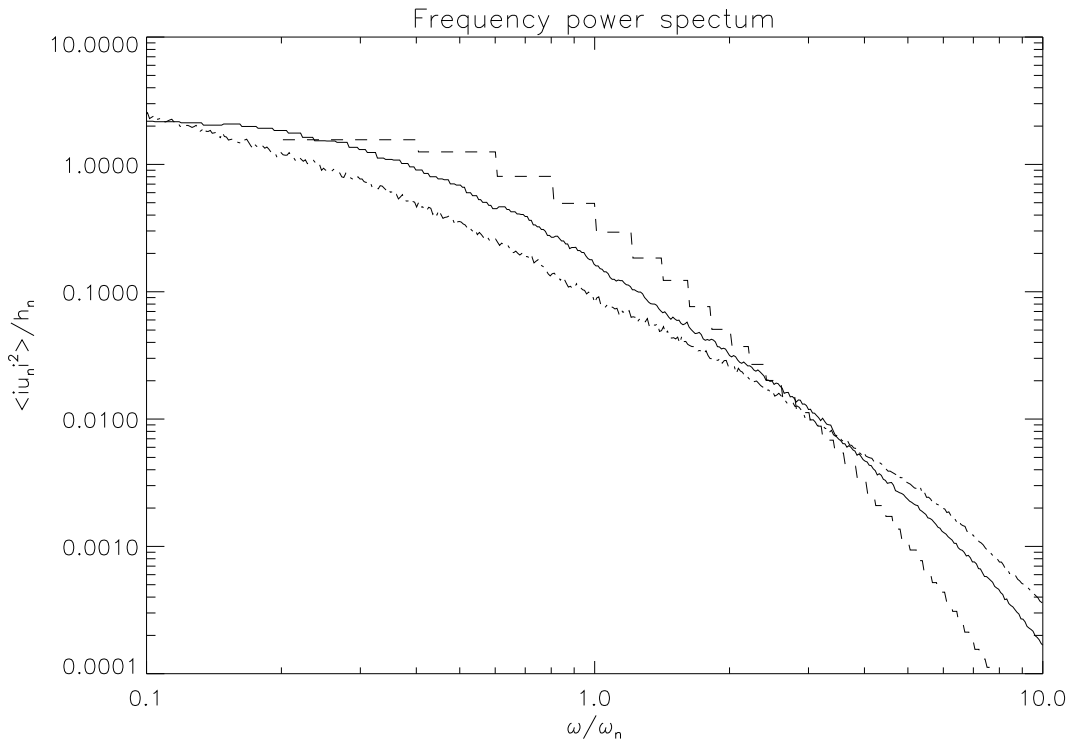

Figure 1: Frequency power spectra for three shells 12(dashed), 22(full), 32(dot dashed) after the rescaling given in text. The intersection point is an artifact of the rescaling and not a real feature of the data. Although there is a superficial similarity, the three functions depicted in this figure differ significantly. We measured the difference by calculating the ratios $\langle\omega^n\rangle/\langle\omega^2\rangle^{n/2}$. These ratios deviate significantly between shells indicating that the functions do not collapse onto a universal scaling function.

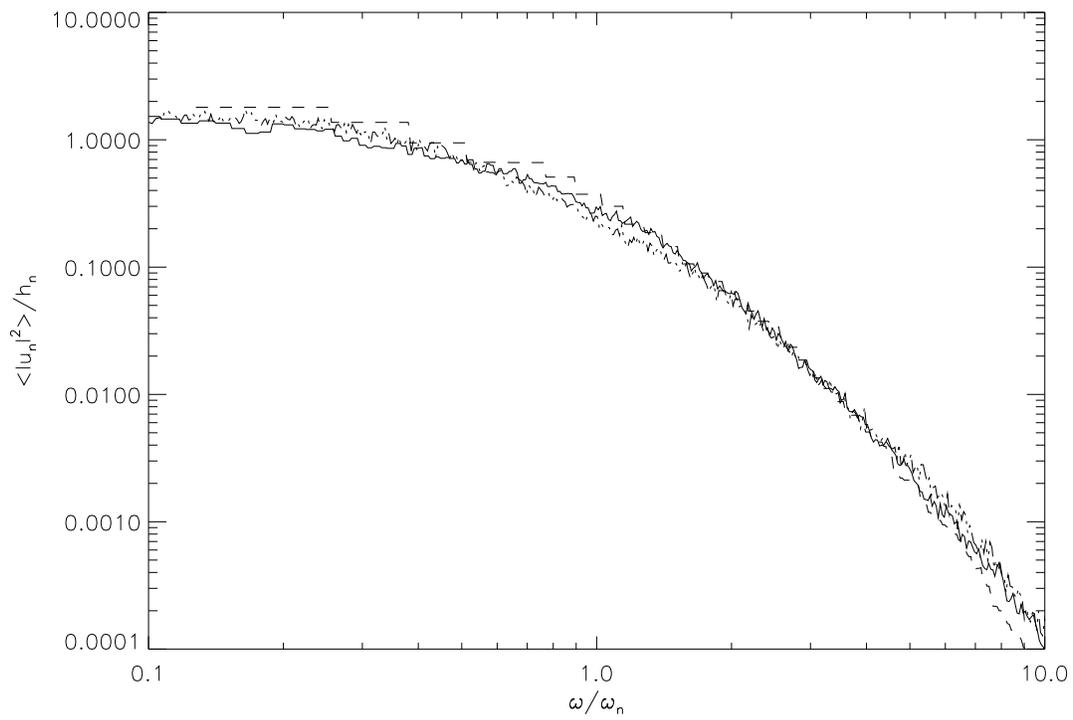

Figure 2: Same as figure 1 for the modified amodel. Note the significant improvement in the fit between the rescaled power spectra.